\documentclass[twocolumn]{aastex62}
\usepackage{apjfonts}

\newcommand{\cnjwl}{$cn_{\rm JWL}$}

\newcommand{\cnwave}{$\lambda$3883}
\newcommand{\scn}{CN(3839)}
\newcommand{\ds}{$\delta$CN(3839)}
\newcommand{\solarmass}{$M_\sun$}
\newcommand{\gaia}{{\it Gaia}}

\shorttitle{CN Toward Galactic Bulge Field}
\shortauthors{Lee}
\begin{document}

\title{Structure of our Galactic Bulge from CN Measurements}

\author[0000-0002-2122-3030]{Jae-Woo Lee}
\affiliation{Department of Physics and Astronomy, Sejong University,
209 Neungdong-ro, Gwangjin-Gu, Seoul, 05006, Republic of Korea;
jaewoolee@sejong.ac.kr, jaewoolee@sejong.edu}

\begin{abstract}
The double red clumps (DRCs) are now dominantly believed to be the strong observational line of evidence of the so-called X-shaped Galactic bar structures.
Recently, \citet{ywlee18} reported a subtle mean \ds\ difference between the DRCs and suggested a dichotomic picture that can be seen in globular clusters: the faint red clump is the first generation, while the bright red clump corresponds to the second generation (SG). They argued that the magnitude difference between the DRCs is due to different stellar populations, and is not due to the geometric difference between the DRCs. Our reanalysis shows that their data do not appear to support the idea of the multiple population-induced DRCs in our Galactic bulge.
We perform fully empirical Monte Carlo simulations and find that the shape of the \ds\ distributions is the most stringent evidence to pursue.
Our results strongly suggest that the CN distributions toward the Galactic bulge are qualitatively consistent with the X-shaped Galactic bulge with a minor fraction of the SG of about 2 -- 3\%.
\end{abstract}


\section{Introduction}
The CN molecules have played some important roles in stellar astrophysics for more than one hundred years \citep{lindblad}, because usually very strong CN absorption strengths can easily be detectable with great accuracy not only in low-resolution spectroscopy but also in photometry \citep[see, e.g.,][]{ddo,lee17,lee18,lee19a}. 
In recent years, the study of CN molecular bands, especially in the blue where the distinctive band features exist, regained its popularity since the CN measurements appear to provide promptly pivotal information on the multiple populations (MPs) in globular clusters (GCs).

However, the understanding of the line formation of  molecular bands may not be as simple as it might seem. For example, the rate of CN molecule formation depends on various factors:
(1) The available carbon and nitrogen atoms that will be involved in the CN molecule formation depend on abundances of other molecules, such as, NH, CH, C$_2$, and CO \citep[see, e.g.,][]{tsuji73}. 
These diatomic molecules have different dissociation energies and therefore they experience different degrees of the luminosity effect and temperature effect.
Similar to other molecules, CN also suffers from luminosity and temperature effects to a rather serious degree \citep[see, e.g.,][]{gray09}.
(2) The CN band strengths can also be affected by the metallicity effect, in particular through the degree of the formation rate of the negative hydrogen ion, which is the dominant continuum opacity source in cool stars \citep[see, e.g.,][]{suntzeff81,lee15}.
(3) The $^{12}$C/$^{13}$C ratio can affect the CN band strengths, in particular the CN band at \cnwave\ \citep[see, e.g.,][]{briley89}.
(4) Finally, the choice of the continuum sidebands for the CN strength measurements could be a matter of potential problem \citep{lee19b}.
Therefore, any low-resolution spectroscopic or photometric studies of the molecular band features should be performed with great caution.

In a rather simple stellar population, such as stars in GCs or open clusters, some of the difficulties mentioned above can be well controlled.
In this context, for example, \citet{norris81} devised an ingenious index, \ds, to delineate the relative CN contents in a GC.\footnote{The \scn\ index used in this Letter is not exactly the same as that defined by \citet{norris81}, $S$(3839). The \scn\ index is defined by \citet{harbeck03} in their study of the main-sequence stars to avoid the contamination from the hydrogen Balmer lines, which become stronger in dwarf stars as with the Stark effect. Therefore the bandwidth and the continuum sideband of the \scn\ are different from those of $S$(3839), and it is very unfortunate that any direct comparisons between these two indices should be avoided.}
Basically, they defined a lower envelope in a plot of the \scn\ versus $V$ magnitude and then they derived the relative \scn\ offset values from the baseline at a given magnitude to estimate the relative CN contents. This approach can overcome the luminosity and temperature effects to some degree without difficulty.
Through this procedure, they constructed a concrete picture of the bimodal CN distributions in GC red giant branch (RGB) stars. 

It should be emphasized that this approach must be applied to a simple population with similar surface gravity and effective temperature.
In heterogeneous systems, the magic of \ds\ is no longer effective \citep[see, e.g.,][]{lee15}.
At the same time, using the baseline, not the linear fit to the data, can be very critical when comparing relative CN contents with different luminosities or effective temperatures.
In spite of its objectiveness, the results from the linear fit can be vulnerable to the incompleteness of the sample, which we will discuss later. 
Therefore, the \ds\ index can be a versatile but deceptive tool.

In our work, we reanalyze the CN measurements toward the Galactic bulge field by \citet[][hereafter YWL18]{ywlee18}, and find that their results are peculiarly biased.
In our work, we will focus on the \scn\ and \ds.
The CN(4142) and CH(4300) indices tend to not show clear bifurcation as the \scn\ and \ds\ do between MPs in GCs, and the former two indices may have potential problems with the continuum sideband assignments \citep{lee19b}.

\section{Notes on Y.-W. Lee et al. (2018)}\label{s:ywlee}
Before we proceed further, we would like to discuss four points on the results presented by YWL18.

First, in Figure~4 of YWL18, it is clear that the range of the \scn\ dispersion of the bright RGB (bRGB) stars with $K_S \lesssim$ 12.1 mag looks much smaller than those of other groups. 
The standard deviation of the bRGB is $\sigma$[\scn] = 0.15, while $\sigma$[\scn] = 0.21, 0.24, and 0.25 for the bright red clump (bRC), faint red clump (fRC), and faint RGB (fRGB), respectively.  We performed randomization tests to see if the CN distribution of the bRGB is the same as other groups, and we found very low probabilities of the bRGB being drawn from the empirical distributions of bRC, fRC, and fRGB, 5.74 $\pm$ 0.07\%, 0.66 $\pm$ 0.02\%, and 0.64 $\pm$ 0.03\%, respectively.\footnote{Our randomization tests for the HK$^\prime$, whose science band lies right next to the CN band at \cnwave, show that the probabilities of the bRGB being drawn from the bRC, fRC, and fRGB are 66.43 $\pm$ 0.04\%, 69.82 $\pm$ 0.15\%, and 38.49 $\pm$ 0.16\%, respectively, suggesting that these four groups are most likely similar in the HK$^\prime$ domain.} 
\emph {From a statistical perspective, it can be said that the bRGB is a totally alien population.}
In addition, what is more difficult to understand is that the locations of the bRGB stars tend to occupy the CN-weak side of the plot.
As we will show later, the standard deviations from each magnitude bin should be similar, if they were drawn from unbiased parent distributions. 
Furthermore, if these bRGB stars were selected from an unbiased sample and they were really bright RGB stars with a mixture of the first generation (FG) and the second generation (SG) of stars as can be found in GCs, bf the $\sigma$[\scn] of the bRGB group is expected to be slightly larger than those of other groups due to the well-known luminosity effect \citep[see, e.g., Figure~16 of][]{lee15}.

Second, YWL18 adopted a linear fit to remove the luminosity effect assuming that all their sample stars are homogeneous and are located at the same distance from us.
As we already mentioned, using the lower envelope baseline is a more fail-free approach. By doing that, the CN difference between the bRC and fRC found by YWL18 can be naturally  reduced or even erased and, as a consequence, a null \ds\ gradient between the bRC and fRC can be established.

\begin{figure*}
\epsscale{1.1}
\figurenum{1}
\plotone{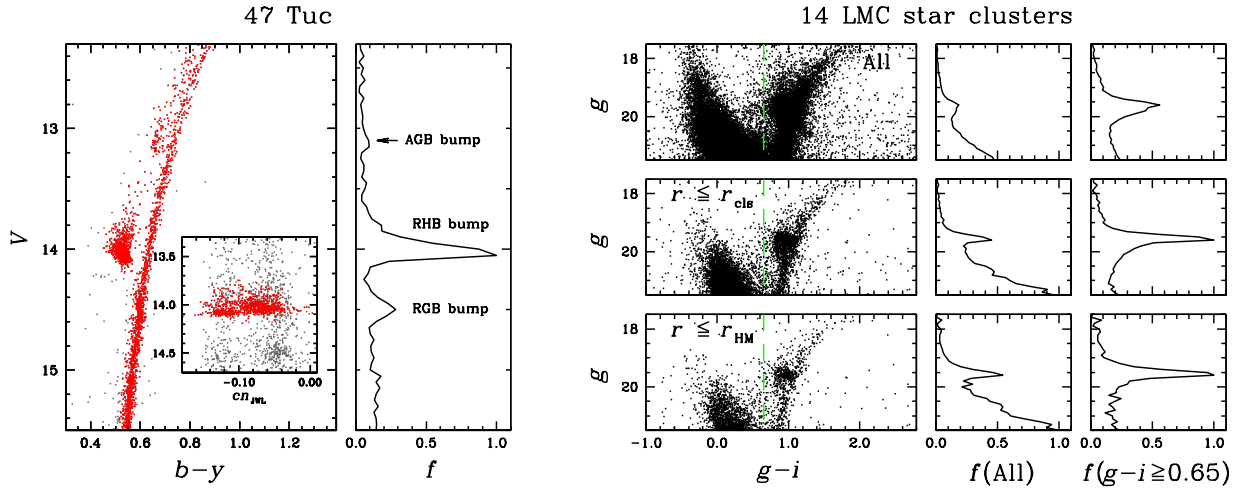}
\caption{
Left top panels: the CMD and the LF for 47 Tuc.
The \cnjwl\ versus $V$ CMD for RHB (red dots) and RGB (gray dots) stars in 47~Tuc is shown in the inset. 
Right panels: CMDs and LFs for 14 LMC star clusters. $r_{\rm cls}$ and $r_{\rm HM}$
denote the cluster and the half-mass radii, while $f(g-i\geq0.65)$ denotes the normalized number of stars with $(g-i)\geq0.65$ mag.
}\label{fig:n104lf}
\end{figure*}

Third, due to the presence of the RGB bump (RGBB) and the increasing stellar number density with magnitude (e.g., see Figure~\ref{fig:n104lf}), the fRC can contain a considerable amount of the SG in a picture of the MP-induced double red clumps (DRCs).
As we will show later, the DRCs in a single bar structure with the empirical luminosity function (LF) of 47~Tuc, the number ratios between the FG and SG, are about 16:84 and 70:30 for bRC and fRC, respectively, which weaken the argument raised by YWL18.

Finally, as we will show later, the presence of the mirror-image asymmetric \ds\ distributions is a more stringent observational line of evidence of the MPs as the origin of the DRCs.
A subtle change in the mean \ds\ can easily be sneaked depending on the adopted slope of the fitted line, especially in heterogeneous stellar systems. 
In sharp contrast, the shape of the \ds\ distribution is less vulnerable to such artifacts.
It is fair to mention that the asymmetric distributions were also hinted at YWL18 in a weak form in their Figure~3, but their results failed to show it.

\section{Monte Carlo Simulations for the DRCs: Symmetric, Skewed, and Asymmetric Mirror-Image Distributions Are Involved}
\subsection{The Red Clump Frequencies of 47~Tuc and the Intermediate-age Large Magellanic Cloud clusters: A Sanity Check}
In order to understand the behavior of the CN distribution between the DRCs with different groups of stars, we constructed fully empirical models to perform Monte Carlo simulations using our high-precision multicolor photometry of 47~Tuc. 

In Figure~\ref{fig:n104lf}, we show our CMD and LF of the bright stars in the metal-rich ([Fe/H] = $-$0.77 dex; \citealt{carretta09}) GC 47~Tuc, where the off-cluster field stars were removed using the Gaia DR2 proper motion study \citep{gaiadr2} and using our multicolor photometry \citep[see, e.g.,][for the versatility of multicolor photometry]{lee15}.
Note that this result is from our work of an 1\degr\ $\times$ 1\degr\ photometric study, aimed at revealing the MPs of the cluster, and will be published in a forthcoming paper.
Our LF of the cluster shows the pronounced RHB bump (RHBB) and RGBB as well. 

47~Tuc is slightly more metal poor than the mean metallicity of the RGB stars in the high galactic latitude bulge field, ($l$, $b$) $\approx$ (1\degr, $-$8\fdg5), that YWL18 studied. \citet{johnson11} performed a high-resolution spectroscopic study of the same field, and they obtained metallicities for 61 RGB stars, finding $-$1.5 $<$ [Fe/H] $<$ +0.3, with a median of about $-$0.4 dex, slightly more metal rich than 47~Tuc.

In order to understand the dependency of the RHB and RC on metallicity, we examined the RC frequency of 14 intermediate-age star clusters with [Fe/H] $\approx-$ 0.4 dex in the Large Magellanic Cloud (LMC) by \cite{piatti14}. 
As shown in Figure~\ref{fig:n104lf}, the LMC clusters also show very pronounced RC bumps (RCBs).
We calculated the number of stars in the RHBB and in the RCBs and then obtained the RHB and RC frequencies by normalizing them with the number of bright and faint RGB stars (1 mag brighter and fainter than the RGBB and RCBs, respectively) within 1 mag bins.
We show our results in Table~\ref{tab:freq} and the RHB frequency of 47~Tuc appears to be in excellent agreement with those of the LMC clusters and, therefore, we conclude that 47~Tuc's RHB can be a good surrogate RC in our Galactic bulge.

\begin{deluxetable}{lcc}
\tablenum{1}
\tablecaption{The observed RHB and RC frequencies per unit magnitude\label{tab:freq}}
\tablewidth{0pc}
\tablehead{
\multicolumn{1}{c}{Obj} &
\multicolumn{1}{c}{$n$(RC)/$n$(bRGB)} &
\multicolumn{1}{c}{$n$(RC)/$n$(fRGB)}
}
\startdata
47 Tuc (RHB) &  6.37  & 2.40   \\
LMC (All $\mid g-i \geq$ 0.65)       &  6.30  & 1.91   \\
LMC ($r_{\rm cls}\mid g-i \geq$ 0.65)       &  7.56  & 2.54   \\
LMC ($r_{\rm HM}\mid g-i \geq$ 0.65)       &  6.40  & 2.55   \\
\enddata 
\end{deluxetable}

\begin{figure*}
\epsscale{1.1}
\figurenum{2}
\plotone{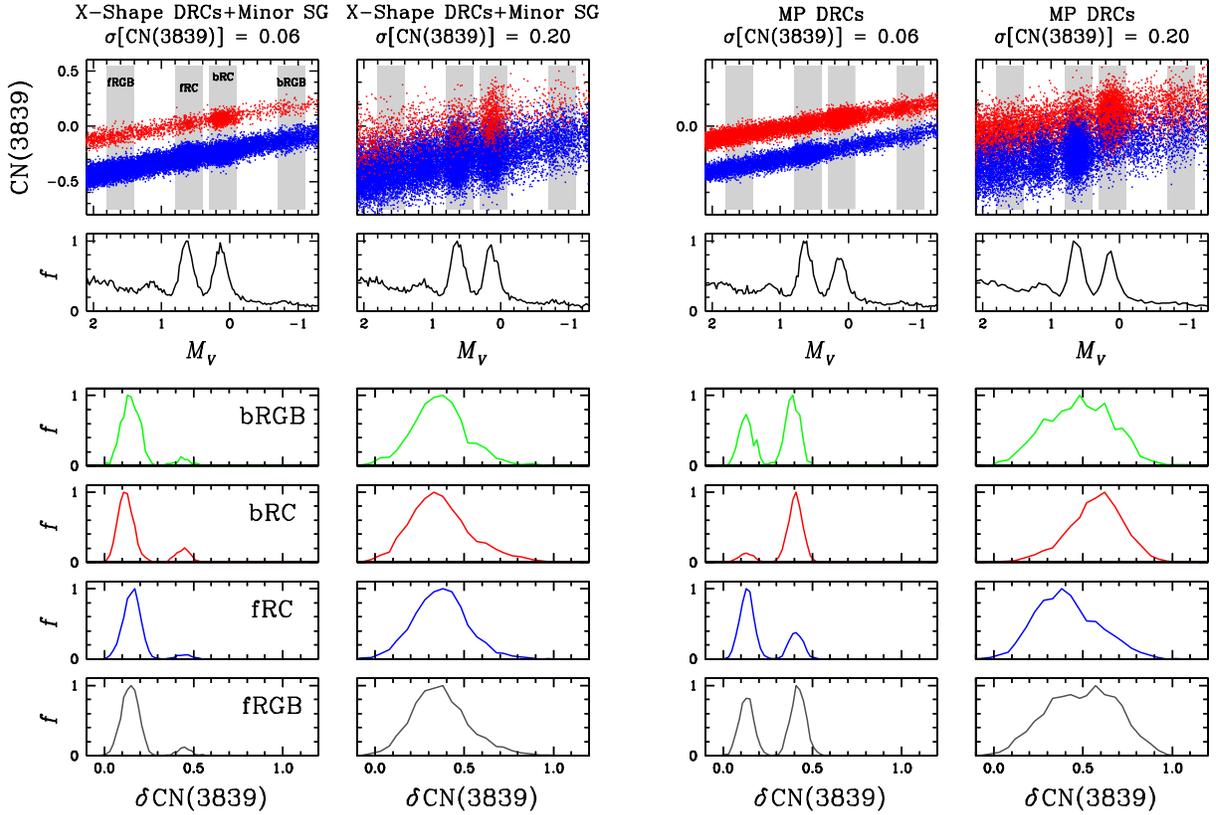}
\caption{
Plots of the synthetic \scn\ versus $M_V$, LFs, and the \scn\ distributions of the bRGB (green), bRC (red), fRC (blue), and fRGB (dark gray).
}\label{fig:n104mc}
\end{figure*}

\subsection{Ingredients for Models}
The detailed underlying basic schemes are already given in our previous work \citep[see][]{lee15}, and here we briefly discussed observational ingredients of our models.
We adopted our 47~Tuc's LF as shown in Figure~\ref{fig:n104lf}, where our \cnjwl\ of the RHB stars are superposed onto the RGB.
In our previous study, we showed that our \cnjwl\ index is a very accurate photometric measure of \scn\ and therefore our results suggest that the \scn\ of the RHB should be very similar to that of the RGB stars with a similar magnitude \citep{lee17,lee18,lee19a}.
We extracted the CN-weak (the first generation, FG) and CN-strong (the second generation, SG) sequences of NGC~362 from \citet{lim16}, who employed the same \scn\ index that YWL18 adopted, and we derived the fiducial sequences for both populations. Then we calculated the scatters of individual stars around the fiducial sequences, finding $\sigma$[\scn] $\approx$ 0.06 for both populations with no metallicity spread. As we mentioned above, due to the presence of a large metallicity spread in the Galactic bulge field, we also adopt $\sigma$[\scn] $\approx$ 0.20 for a heterogeneous population system in our simulations \citep[e.g., see][]{lee15}.

Our previous studies suggested that the LFs of individual populations in GCs can be slightly different \citep[see, e.g.,][]{lee15,lee17,lee18}.
However, we emphasize that slightly different LFs between MPs in GCs do not affect our results presented here.
As our results will show, the populational number ratio is thought to be a more important ingredient in our models.

\begin{figure}
\epsscale{1.1}
\figurenum{3}
\plotone{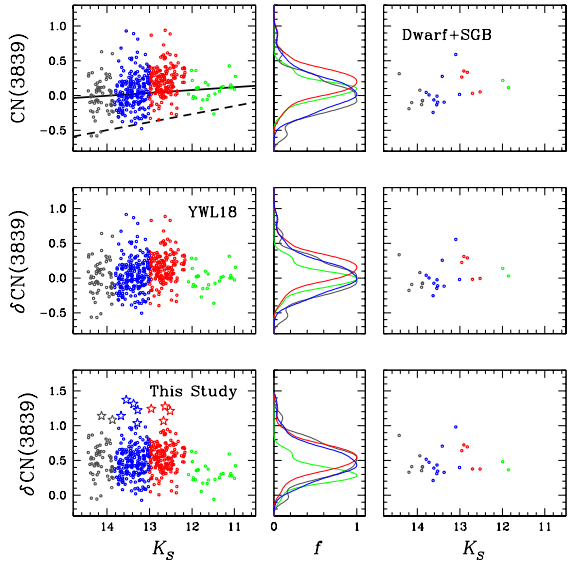}
\caption{
(Left panels) Plots of the \scn\ and \ds\ against $K_S$. The solid line denotes the fitted line by YWL18, while the dashed line denotes our baseline.
In the bottom panel, a group of stars shown with star marks is most likely the SG in our Galactic bulge.
(Middle panels) Histograms for the \scn\ distributions. The gray, blue, red, and green solid lines denote the \scn\ and \ds\ distributions of the fRGB, fRC, bRC, and bRGB, respectively.
Note that the bRGB does not agree with those from other groups, a strong evidence that the bRGB sample by YWL18 is heavily biased.
(Right panels) Stars classified as main-sequences or sub-giants from the second \gaia\ date release.
}\label{fig:dcn}
\end{figure}

\subsection{Single RC Population in an X-shaped Bulge+Minor SG Population}
First, we developed a model for the single RC population in an X-shaped Galactic bulge, which is a dominantly accepted picture for our Galactic bulge \citep[see, e.g.][]{mcwilliam10,wegg13}.
In our model, we allocated the even number of stars to the individual bar branches and we assumed that they are all composed of the FG population. 
In addition to the DRCs, we also added a minor SG population (an additional 10\% of the total number of stars). 

Figure~\ref{fig:n104mc} shows plots of the \scn\ versus $M_V$, LFs, and \ds\ distributions for two sets of models with $\sigma$[\scn] = 0.06 and 0.20.
To calculate \ds, we adopted the scheme that \citet{norris81} devised.
The influence of the RGBBs is noticeable in LFs: the RGBB of the fRC population is clearly visible at $M_V$ $\approx$ 1.2 mag, while the RGBB of the bRC population is superposed onto the fRC population. 
We also calculated the histograms for the bRGB (1.0 mag brighter than the bRC) and the fRGB (1.0 mag fainter than the fRC), which can serve as references.

The \ds\ value at the peak of the bRC distribution is slightly smaller than that of the fRC, but we strongly believe that this small difference should not be considered as conclusive evidence. 
Instead, the shape of the \ds\ distribution, which is almost independent of the adopted baseline during the \ds\ calculations, provides a more informative and, perhaps, the most practical probe to explore the existence of the SG population in our Galactic bulge.
Without the minor SG population, the shapes of all \ds\ distributions are almost symmetric: not only the bRC and fRC but also the bRGB and fRGB.
When some fractions of the minor SG population are included, the shapes of the \ds\ distribution show a weak secondary peak at larger \ds\ regimes.

In the case of $\sigma$[\scn] = 0.20, which is for the heterogeneous stellar populations with a metallicity spread, the inclusion of the minor SG population makes all the \ds\ distributions skewed to larger \ds, not to smaller \ds. 

\subsection{Double RC Populations in a Single Bar}
To examine the idea proposed by YLW18, we constructed models for a single Galactic bar with MPs, where the fRC corresponds to the FG population, while the bRC corresponds to the SG population. Our results are shown in Figure~\ref{fig:n104mc}.

In the model with a GC-like dispersion, i.e., $\sigma$[\scn] = 0.06, the discrete double sequences are eminent for all magnitude regimes. Moreover, the shape of the \ds\ distributions of the bRC and fRC show a mirror-image characteristic, leaving a distinctive and the most profound observable footprint, which has not been observed so far. For both RGB regimes, the relative frequencies of the peaks with a large \ds\ are slightly larger than those of the peaks with a small \ds, due to slight differences in the evolutionary speed with luminosity.

In the case of $\sigma$[\scn] = 0.20, the \ds\ distributions of the bRC and fRC are asymmetric and show a mirror-image characteristic, while those of the RGB regimes show near-symmetric distributions, which show very different characteristics from those from a single RC population in an X-shaped bulge.

\subsection{Comparisons with YWL18}
In Figure~\ref{fig:dcn}, we show plots of \scn\ and \ds\ against $K_S$ and histograms for each group of stars that YWL18 studied.
As we mentioned earlier, the slope from the fitted line adopted by YWL18 is slightly different from the slope of the baseline that is conventionally used.
In the figure, it is very clear that the histograms for the bRGB are in total disagreement with those of other groups, suggesting that the bRGB sample must have been heavily biased.
None of our simulations can explain this strange behavior of the bRGB stars.
This is very critical in comparing \ds\ values from different magnitudes as the way YWL18 did, as it is most likely that they relied on the biased reference fitted line to derive their \ds.
It is strongly believed that the CN gradient between the bRC and fRC reported by YWL18 is not real but is mainly due to the incorrect assignment of the reference line.

We emphasize that the observed \ds\ distributions other than the bRGB are qualitatively very consistent with our results from the single RC population in an X-shaped Galactic bulge with a minor SG population.
If this is the case, about 11 stars ($\approx$ 2.4\%) with \ds\ $\gtrsim$ 1.0 in our analysis could be the SG of stars in our Galactic bulge.

Stellar parameters from the second \gaia\ data release can provide a wonderful opportunity to assure the evolutionary stage of YWL18's sample.  We cross-matched the coordinates of the YWL18's sample and we were able to identify 25 stars.
Unfortunately, all of them were turned out to be either the dwarf or subgiant based on the radius and luminosity estimates \citep{apsis}.

In Figure~\ref{fig:dcn}, we also show plots of the \scn\ and \ds\ versus $K_S$ for the known dwarfs and subgiants in YWL18's sample. We calculated the \ds\ difference for the dwarfs and subgiants, finding that $\Delta$\ds\ = 0.155 $\pm$ 0.091 in YWL18's scale, in the sense that the \ds\ value of the dwarfs and subgiants included in the bRC group is larger than that in the fRC.
Furthermore, $\Delta$\ds\ from the dwarfs and subgiants is consistent with that between the bRC and fRC found by YWL18, $\Delta$\ds\ = 0.125 $\pm$ 0.023. This also indicates that the \ds\ gradient argued by YWL18 is most likely an artifact.
In sharp contrast, from our analysis the $\Delta$\ds\ between the two groups of dwarfs and subgiants is 0.100 $\pm$ 0.092, and the gradient in the \ds\ becomes almost null, which is very natural to expect.

Our exercises presented here strongly suggest that the \ds\ gradient between the two RCs reported by YWL18 is mainly due to two reasons: 
(1) the use of an incorrect assignment of the reference line in deriving \ds, and
(2) the inclusion of the nonnegligible fraction of the misrepresentative samples.
Therefore, one can naturally argue that the results presented by YWL18 do not support the idea of the MP-induced DRCs in our Galactic bulge.

\section{Summary and Discussion}
In order to understand what stellar populations constitute the X-shaped Galactic bulge, we performed fully empirical Monte Carlo simulations.
In sharp contrast to YWL18, who argued a small \ds\ difference between the DRCs as conclusive observational evidence, our study highlighted the importance of examining the shape of the \ds\ distributions to understand underlying stellar populations.
Our results strongly suggested that the CN measurements by YWL18 are qualitatively more consistent with the X-shape Galactic bulge with a minor SG fraction of about 2 -- 3 \% \citep[see, e.g.,][]{martell11}.
We also showed the inclusion of a nonnegligible fraction of dwarfs and subgiants in the sample by YWL18.
Furthermore, their bRGB sample appeared to be severely and quizzically  biased, insufficient for being a reference in deriving the \ds\ index.

If about an half of the Galactic bulge stars are formerly GC SG stars that are now dissolved, as YWL18 suggested, then one can anticipate the clear and present difficulty of the absence of the reservoir of the dissolved FG population, which supplied an enormous amount of lighter elements to make the SG population that is now observed as the bRC, as YWL18 claimed.
The total mass of our Galactic bulge is about 2 $\times$ 10$^{10}$\solarmass, which is about 1000 time more massive than the total mass of the Galactic GCs
\citep[see, e.g.,][]{valenti16}.
One of the unsolved problems in the formation of GCs with MPs is the so-called mass-budget problem.
To explain the chemical evolution of GCs, at least about 10 to 100 times more FG stars were required in the past to form the SG population than can be found in GCs, and most of the FG populations in Galactic GC systems must have been lost during the early phases of the GC evolution. The main body of our Galaxy must have been the reservoir of the GC FG stars that are now dissolved.
Then where did the FG of the bRC go? 
If about the half of the Galactic bulge is the SG, then more than about 10$^{11}$ -- 10$^{12}$\solarmass\ of the formerly GC FG stars now dissolved in our Galaxy, including the Galactic bulge, which will make the populational number ratio of 10 -- 100:1 between the FG and SG.
In our Galactic bulge, however, the populational number ratio should be about 1:1 between the FG and SG based on the observed number ratio between the fRC and bRC, and the number simply does not add up in the dichotomic population picture by YWL18.

We strongly believe that, for example, sodium abundance measurements from high-resolution spectroscopy with good estimates of the stellar parameters will shed more light on the MPs in our Galactic bulge in the future.

\acknowledgements
J.-W.L. sincerely thanks the Lord for having extremely bad weather conditions and fatal instrument failures during his observing run in 2019 February, which enabled him to conceive the idea and complete this work in a very short period of time. 
The anonymous referee is thanked for useful comments.
Financial support from the Basic Science Research Program (grant No. 2016-R1A2B4014741) through the National Research Foundation of Korea (NRF) funded by the Korea government (MSIP) is acknowledged.

\end{document}